\documentclass[pdflatex,sn-nature]{sn-jnl}
\usepackage{float}
\usepackage{mathrsfs}
\usepackage{graphicx}%
\usepackage{multirow}%
\usepackage{amsmath,amssymb,amsfonts}%
\usepackage{amsthm}%
\usepackage{mathrsfs}%
\usepackage[title]{appendix}%
\usepackage{xcolor}%
\usepackage{textcomp}%
\usepackage{manyfoot}%
\usepackage{booktabs}%
\usepackage{float}
\usepackage{gensymb}
\theoremstyle{thmstyleone}%
\usepackage[T1]{fontenc}

\theoremstyle{thmstyletwo}%

\theoremstyle{thmstylethree}%

\begin{document}

\title[Article Title]{Localized Excitonic Emission in Wafer-Scale MOCVD-Grown GaSe 2D Nanosheets for Classical and Non-Classical Light Sources}

\author[1]{\fnm{Bhabani Sankar} \sur{Sahoo}}
\equalcont{These authors contributed equally to this work.}
\author[2,3]{\fnm{Nils Fritjof} \sur{Langlotz}}
\equalcont{These authors contributed equally to this work.}

\author[1]{\fnm{Shachi} \sur{Machchhar}}
\author[1]{\fnm{Kartik} \sur{Gaur}}
\author[2,3]{\fnm{Robin} \sur{Günkel}}
\author[2,3]{\fnm{Max} \sur{Bergmann}}
\author[2,3]{\fnm{Naghmeh} \sur{Ghadghooni}}
\author[1]{\fnm{Aris} \sur{Koulas-Simos}}
\author[2,3]{\fnm{Jürgen} \sur{Belz}}
\author[1]{\fnm{Chirag Chandrakant} \sur{Palekar}}
\author[1,4]{\fnm{Maximilian} \sur{Ries}}
\author[2,3]{\fnm{Kerstin} \sur{Volz}}
\author[1]{\fnm{Stephan} \sur{Reitzenstein}}
\author*[1]{\fnm{Imad} \sur{Limame}}\email{imad.limame@tu-berlin.de}

\affil[1]{\orgdiv{Institut für Physik und Astronomie}, \orgname{Technische Universit\"at Berlin}, \orgaddress{\street{Hardenbergstrasse 36}, \city{Berlin}, \postcode{10623}, \country{Germany}}}

\affil[2]{\orgdiv{mar.quest | Marburg Center for Quantum Materials and Sustainable Technologies}, \orgname{Philipps-Universität Marburg}, \orgaddress{\street{Hans Meerwein Str. 6}, \city{Marburg}, \postcode{35032}, \country{Germany}}}

\affil[3]{\orgdiv{Department of Physics, Structure and Technology Research Lab}, \orgname{Philipps-Universität Marburg}, \orgaddress{\street{Hans Meerwein Str.~6}, \city{Marburg}, \postcode{35032}, \country{Germany}}}

\affil[4]{\orgdiv{Chemical Analysis Division}, \orgname{Thermo Fisher Scientific GmbH}, \orgaddress{\street{Im Steingrund~4-6}, \city{Dreieich}, \postcode{63303}, \country{Germany}}}

\abstract{Wafer-scale growth of two-dimensional semiconductors remains a key challenge for their integration into photonic technologies. While most studies of two-dimensional semiconductors have focused on transition metal dichalcogenides and their scalable fabrication, comparatively little attention has been given to III–VI monochalcogenides. Here, we report wafer-scale growth of gallium selenide (GaSe) by metal–organic chemical vapor deposition (MOCVD) and investigate its structural and optical properties for visible-range classical and quantum light emission.Two samples with thicknesses ranging from a few monolayers to several micrometers, controlled via the growth time, were investigated. The 30-minute grown sample yields intense, broad photoluminescence spanning 1.7-2.0$\,$eV, whereas the thinner 3-minute sample exhibits discrete narrow emission lines and single-photon emission with $(g^{(2)}(0) = 0.15 \pm 0.10)$. Remarkably, cathodoluminescence mapping reveals pronounced spatial localization of both spectrally narrow and broad emission centers. Together with temperature-dependent power-law analysis and Raman mapping, our results indicate defect-induced emission rather than intrinsic excitonic recombination. These findings establish wafer-scale MOCVD grown 2D GaSe as a platform for classical and non-classical light sources and highlight defect-engineered localization as a route toward scalable quantum photonics.}




\maketitle

\section{Introduction}\label{sec1}

The discovery of graphene and its remarkable properties triggered a wave of research in the broader family of two-dimensional (2D) materials~\cite{gb_graphene, Gerstner2010}. In particular, transition metal dichalcogenides (TMDCs) have been extensively studied due to their tunable layer-thickness-dependent band structures, where indirect-to-direct bandgap transitions occur in the monolayer limit~\cite{Bhimanapati.2015, XIA20171}. Monolayer TMDCs have furthermore demonstrated excellent performance in nanophotonic applications, including single-photon emission \cite{He2015, Tonndorf2015}, strong light-matter interaction \cite{Liu2015} and low-threshold high-$\beta$ lasing \cite{Wu2015, Barth2024, Koulas2024}, driven by strong excitonic effects. While TMDCs have dominated research in two-dimensional semiconductors, the broader family of layered materials encompasses a variety of systems with distinct electronic and optical properties that remain comparatively underexplored. Among these, post-transition metal chalcogenides (PTMCs) such as gallium selenide (GaSe) have recently emerged as a promising class of layered materials, offering strong nonlinear optical responses for second-harmonic generation~\cite{Leisgang.2018}, high carrier mobility~\cite{derguenkler, Hilse.2025} relevant for next-generation low-power electronics~\cite{Liu_zhuojun}, thickness-dependent optical emission~\cite{Zallo.2023, Mameyer.2024, Jung.2015}, and compatibility with silicon-based growth platforms~\cite{Zallo.2023}. GaSe is a layered semiconductor composed of Se–Ga–Ga–Se tetralayers with weak van der Waals bonding between adjacent layers. In multilayer form, GaSe exhibits a direct bandgap in the range of 1.74--2.0~eV, depending on strain and defects, while in the monolayer limit the bandgap increases to $\sim$3.3~eV~\cite{Diep.2024, Barker2025GaSeElastoOptic, Liu2020}. This pronounced spectral tunability, combined with the material’s strong nonlinear optical response, positions GaSe as a compelling candidate for applications in nano- and quantum photonics, including $\mu$LEDs~\cite{muLED}, lasers~\cite{lasers}, second-harmonic generators~\cite{Zhou.2015}, and single-photon sources~\cite{Rakhlin.2024, Tonndorf.2017}.

Translating the promising properties of 2D GaSe into practical devices critically depends on the availability of scalable and controllable growth approaches. Although significant progress has been made in the study of two-dimensional materials, most experimental work on GaSe has relied on mechanically exfoliated flakes~\cite{1transfer2} or conventional powder-based chemical vapor deposition (CVD)~\cite{cvd2, cvd3}. While exfoliation yields high-quality crystals, it is inherently limited in lateral size, reproducibility, and device integration, making it unsuitable for scalable technologies and real-world applications. Similarly, standard CVD approaches often face challenges in achieving uniform thickness, phase purity, and wafer-scale control~\cite{D1NA00171J}.

In contrast, epitaxial growth techniques provide a pathway toward controlled, large-area synthesis with improved reproducibility and compatibility with existing device architectures~\cite{D5NR02265G}. Among these, metal-organic chemical vapor deposition (MOCVD) is particularly attractive due to its precise control over precursor delivery and growth conditions, enabling uniform films over wafer-scale substrates~\cite{Stringfellow.2004}. Alternative approaches such as molecular beam epitaxy (MBE) and atomic layer deposition (ALD) also offer routes to thin-film growth, but are limited by ultra-high vacuum requirements in the case of MBE~\cite{Zallo.2023} and by the availability of suitable precursors in ALD~\cite{ALD}.

Within this context, GaP/Si substrates represent a promising platform for the epitaxial growth of GaSe, owing to their favorable lattice compatibility and established role in silicon-based optoelectronics for both classical and quantum light sources~\cite{Limame.2024, Limame2026APLPhotonics, 10.1063/1.3624927}. Notably, GaP has already been demonstrated as a viable substrate for the epitaxial growth of other 2D materials, including WSe$^{}_{2}$, HfSe$^{}_{2}$, and TaSe$^{}_{2}$, further underscoring its versatility for integrating layered semiconductors with silicon technology~\cite{mbe_gap_2, mbe_gap_3, mbe_gap_4}. Building on this, in our previous study~\cite{langlotz2025mocvd}, we demonstrated the controlled MOCVD growth of phase-pure 2D GaSe and Ga$_2$Se$_3$ on GaP/Si substrates and established a temperature–composition phase diagram to map their respective stability regimes. 

In this work, we investigate the optical emission properties of wafer-scale MOCVD-grown thin GaSe on GaP/Si, with particular emphasis on excitonic recombination and the emergence of spectrally sharp, spatially localized emission features. These localized emission centers are examined in the context of their linewidth characteristics and potential relevance for classical and non-classical light emission.

To achieve this, GaSe layers were grown on GaP/Si substrates by MOCVD using triethylgallium (TEGa) and diisopropylselenide (DiPSe) precursors, following the deposition of a thin GaP buffer layer formed from TEGa and tertiarybutylphosphine (TBP). Two samples with different GaSe thicknesses ranging from a few monolayers to several micrometers were obtained by varying the growth time to 3 and 30 minutes. The structural properties were investigated using atomic force microscopy (AFM), scanning electron microscopy (SEM), X-ray diffraction (XRD), and Raman spectroscopy to probe surface morphology and crystalline quality. High-resolution transmission electron microscopy (TEM) further provided insight into the epitaxial alignment and the interface structure between the thin GaSe layer and the underlying GaP/Si substrate. Furthermore, we employ high-resolution Raman mapping at room temperature to assess the growth quality and spatial uniformity, simultaneously capturing both the Raman spectral signatures and the photoluminescence (PL) emission. 

Subsequently, the optical properties were characterized by micro-PL ($\mu$PL), time-resolved photoluminescence (TRPL), and cathodoluminescence (CL) spectroscopy, enabling a comprehensive analysis of the spectral characteristics and spatial distribution and recombination dynamics, including exciton lifetimes, of the emission. Finally, the narrow and spatially localized emission lines observed in the 3-minute-grown sample were analyzed using a Hanbury Brown and Twiss (HBT) interferometer, confirming the quantum nature of the emitted light.

\section{Results}\label{sec2}

In this section, we first present the structural characterization of the two epitaxially grown GaSe samples, including AFM, SEM, Raman spectroscopy and mapping, as well as TEM. These measurements provide insight into the surface morphology, crystalline quality, and epitaxial structure of the grown layers. We then discuss the CL mapping results, focusing on the spatial localization of the emission and the observed variations in the spectral response across the samples. Finally, we analyze the optical properties obtained from $\mu$PL measurements, including temperature-dependent PL studies, and present HBT measurements confirming the single-photon emission from a spatially localized emitter in the 3-minute-grown sample.

\subsection{Structural properties of MOCVD-grown GaSe}
\begin{figure}[ht]
 \centering\includegraphics[width=0.9\textwidth]{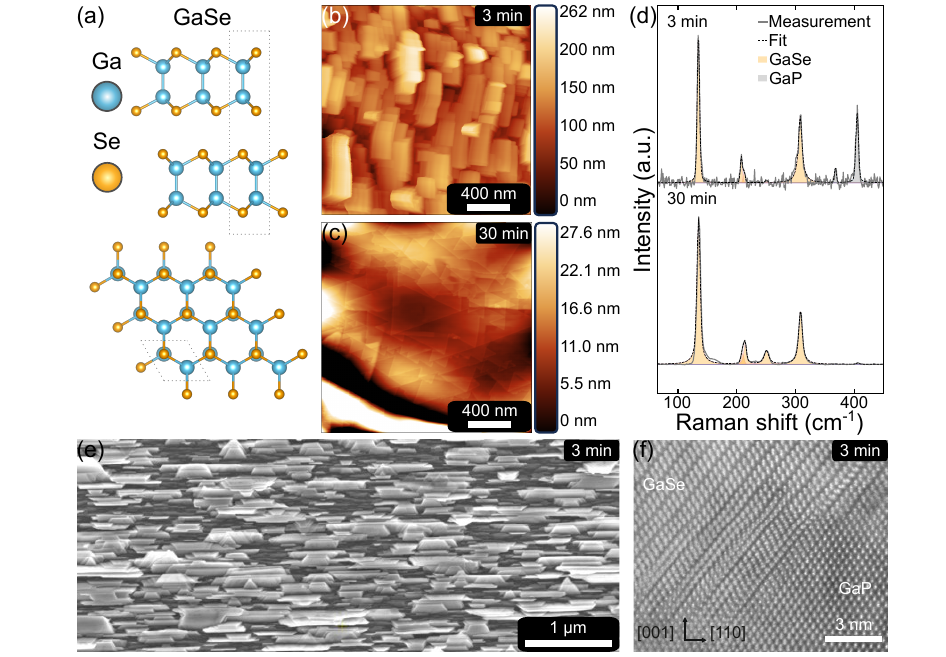}
\caption{\textbf{Structural Analysis of GaSe/GaP/Si.} 
\textbf{a)} Side (top) and top (bottom) views of the GaSe crystal structure, generated using VESTA~\cite{Momma.2011}. 
\textbf{b)} AFM image (2×2 $\mu$m$^2$) showing out-of-plane GaSe nanosheets with a height of up to 262$\,$nm, grown on a GaP/Si substrate after a 3-minute growth duration. The nanosheets exhibit preferential growth along the ⟨111⟩ direction of the GaP substrate.
\textbf{c)} AFM image (2×2 $\mu$m$^2$) showing planar-trigonal GaSe flakes between taller vertical features, obtained after increasing the growth duration to 30$\,$minutes. 
\textbf{d)}  Raman spectra for the samples shown in (b) and (c), exhibiting characteristic vibrational modes of GaSe (highlighted in orange). An additional out-of-plane GaSe mode at $250.6 \pm 1\,$cm$^{-1}_{}$ is observed, in agreement with previous reports~\cite{Grzonka.2021, Diep.2019, Rodriguez.2014, Avdienko.2019, Diep.2024}. Substrate-related GaP peaks (highlighted in grey) are consistent with literature~\cite{langlotz2025mocvd}. The GaP bands appear more pronounced for the thinner sample, while their relative intensity decreases with increasing GaSe thickness due to the dominance of the GaSe signal. Prior to peak fitting, a baseline was estimated and subtracted using the asymmetrically reweighted penalized least squares (arPLS) algorithm.
\textbf{e)} SEM image corresponding to (b), highlighting the sheet-like, out-of-plane growth of GaSe along the ⟨111⟩ direction of the GaP substrate.
\textbf{f)} High-resolution STEM image of the GaP/GaSe interface. The GaP region is visible in the bottom right corner, while the GaSe region appears in the top left. A clear epitaxial relationship between GaP and GaSe is observed at the interface.}

\label{fig1}
\end{figure}

Figure~\ref{fig1}(a) shows the top and side views of the GaSe crystal structure generated using VESTA~\cite{Momma.2011}, highlighting its layered hexagonal arrangement of Se–Ga–Ga–Se tetralayers. AFM imaging of samples grown for a duration of 3-minutes reveals GaSe nanosheets with heights below 262$\,$nm (Figure~\ref{fig1}(b)). These structures exhibit predominantly out-of-plane growth along the ⟨111⟩ directions of the GaP substrate. This behavior is attributed to the surface morphology of GaP, which, due to unintentional surface roughness, exposes (111)-like facets rather than a perfectly flat (100) surface~\cite{langlotz2025mocvd}. The hexagonal symmetry and small lattice mismatch ($<3\,$\%) between 2D GaSe and GaP (111) further support this preferential orientation. These observations indicate that, at short growth times, the GaSe growth is dominated by anisotropic, facet-driven nucleation, resulting in vertically oriented nanosheet formation. In contrast, increasing the growth duration by an order of magnitude to 30 minutes leads to the formation of planar, trigonal GaSe flakes on a predominantly flat GaP (100) surface, appearing between higher out-of-plane features (Figure~\ref{fig1}(c)). This transition suggests a growth-mode evolution toward more laterally extended, energetically favorable configurations with increasing deposition time. SEM imaging of the 3-minute sample (Figure~\ref{fig1}e) confirms the sheet-like, anisotropic morphology of the out-of-plane GaSe structures and corroborates the observed ⟨111⟩-oriented growth.

The Raman spectra of the 3-minute and 30-minute GaSe samples (Figure~\ref{fig1}(d)) exhibit the characteristic vibrational modes of 2D GaSe, with prominent peaks around $\sim$135, 208--210, 212--214, 250, and 308--309~cm$^{-1}$. 
The exact peak positions and associated uncertainties for each sample are summarized in Table~\ref{tab:raman_peaks}, and are consistent with previously reported values~\cite{Grzonka.2021, Diep.2019, Rodriguez.2014, Avdienko.2019, Diep.2024}.

\begin{table}[h]
\caption{Fitted Raman peak positions (cm$^{-1}$) for 3-minute and 30-minute GaSe samples. Uncertainties are $\pm 1\,\mathrm{cm}^{-1}$ unless indicated otherwise.}
\label{tab:raman_peaks}
\begin{tabular*}{\textwidth}{@{\extracolsep\fill}|l|c|c|}
\hline
\textbf{Mode} & \textbf{3 minute sample (cm$^{-1}$)} & \textbf{30 minute sample(cm$^{-1}$)} \\
\hline
A$^{1}_{1g}$      & $134.8$           & $135.0$           \\
E$^{2}_{1g}$     &  $207.9$          & $210.0 \pm 7.3$             \\
E$^{2}_{2g}(TO)$     & $212.7 \pm 1.4$           & $214.3 \pm 3.8$             \\
E$^{1}_{2g}(LO)$     &            & $250.6$                 \\
A$^{2}_{1g}$      & $308.1$           & $308.6$             \\
\hline

\end{tabular*}

\end{table}

The E$^{1}_{2g}$(LO) mode at $\sim$250~cm$^{-1}$ is attributed to an out-of-plane vibrational mode. The spectral region around 210~cm$^{-1}$ was fitted using two components to account for contributions at slightly different frequencies, which have been associated with screw dislocation-driven growth or $\gamma$ to $\gamma^{\prime}$ stacking variations~\cite{Diep.2024, langlotz2025mocvd}. This is consistent with the $\gamma^{\prime}$ stacking configuration observed in the high-resolution STEM analysis discussed below. Prior to peak fitting, a baseline was estimated and subtracted using the asymmetrically reweighted penalized least squares (arPLS) algorithm.

Notably, the A$^{1}_{1g}$ mode at 135$\,$cm$^{-1}_{}$ exhibits a slight broadening (FWHM increase of ($4 \pm 3)\,$cm$^{-1}$) from the 3-minute to the 30-minute sample, indicative of a modest decrease in crystallinity due to increased surface roughness and a higher density of defects arising from the longer growth time. Raman contributions from the GaP substrate (highlighted in orange) are consistent with reported literature values of 367 and 402$\,$cm$^{-1}_{}$~\cite{langlotz2025mocvd}.
While the 30 minute grown sample exhibits a weak shoulder at the A$^{1}_{1g}$ GaSe mode (135 cm$^{-1}$), which could suggest the presence of Ga$_2$Se$_3$, this feature is most likely attributed to degradation or laser-induced decomposition of GaSe~\cite{SMIRI2024112256}. 

Overall, the Raman data point toward predominantly phase-pure GaSe, with no indication of elemental selenium. Importantly, this interpretation is further supported by complementary structural analysis. X-ray diffraction ($2\theta - \omega$ scan, see Figure~S2) reveals only (00X) reflections of GaSe, with no detectable signatures of Ga$_2$Se$_3$. The extracted lattice constant of $(15.85 \pm 0.10)\,$Å is in good agreement with the literature value of $15.949\,$Å~\cite{Barker2025}.

The absence of Ga$_2$Se$_3$ reflections in XRD, despite its weak indication in Raman studies for the 30 minute sample, suggests that the observed feature is not representative of a secondary crystalline phase but rather arises from local, likely laser-induced decomposition effects. Taken together, these findings confirm the high crystalline quality, phase purity, and largely strain-relaxed nature of the epitaxial GaSe layers, while maintaining their intrinsic crystal structure despite the heteroepitaxial integration.

High-resolution scanning transmission electron microscopy (STEM) imaging of the GaSe/GaP interface (Figure~\ref{fig1}(f)) reveals an atomically abrupt interface with a well-defined epitaxial relationship between the two materials. The GaSe region, located in the top left, aligns crystallographically with the underlying GaP substrate, seen in the bottom right. The TEM analysis further indicates a $\gamma^{\prime}_{}$ stacking sequence in the GaSe layers~\cite{Grzonka.2021}, consistent with the additional Raman feature observed near 207--210$\,cm^{-1}_{}$. This atomic-level alignment supports the observed ⟨111⟩-oriented out-of-plane growth and confirms the structural compatibility between GaSe and the faceted (111)-like surfaces of GaP.

Overall, these results demonstrate that the growth duration strongly influences both the morphology and stacking configuration of GaSe, which directly affects the optical properties discussed in the following sections. 

\vspace{0.3em}

\subsection{Spatially resolved Raman and photoluminescence spectroscopy at room temperature}

While the previous Raman analysis focused on globally averaged vibrational properties and structural phase identification, spatially resolved Raman spectroscopy provides further insight into the lateral homogeneity of the GaSe layers and their correlation with the photoluminescence response.

\begin{figure}
 \centering\includegraphics[width=12cm]{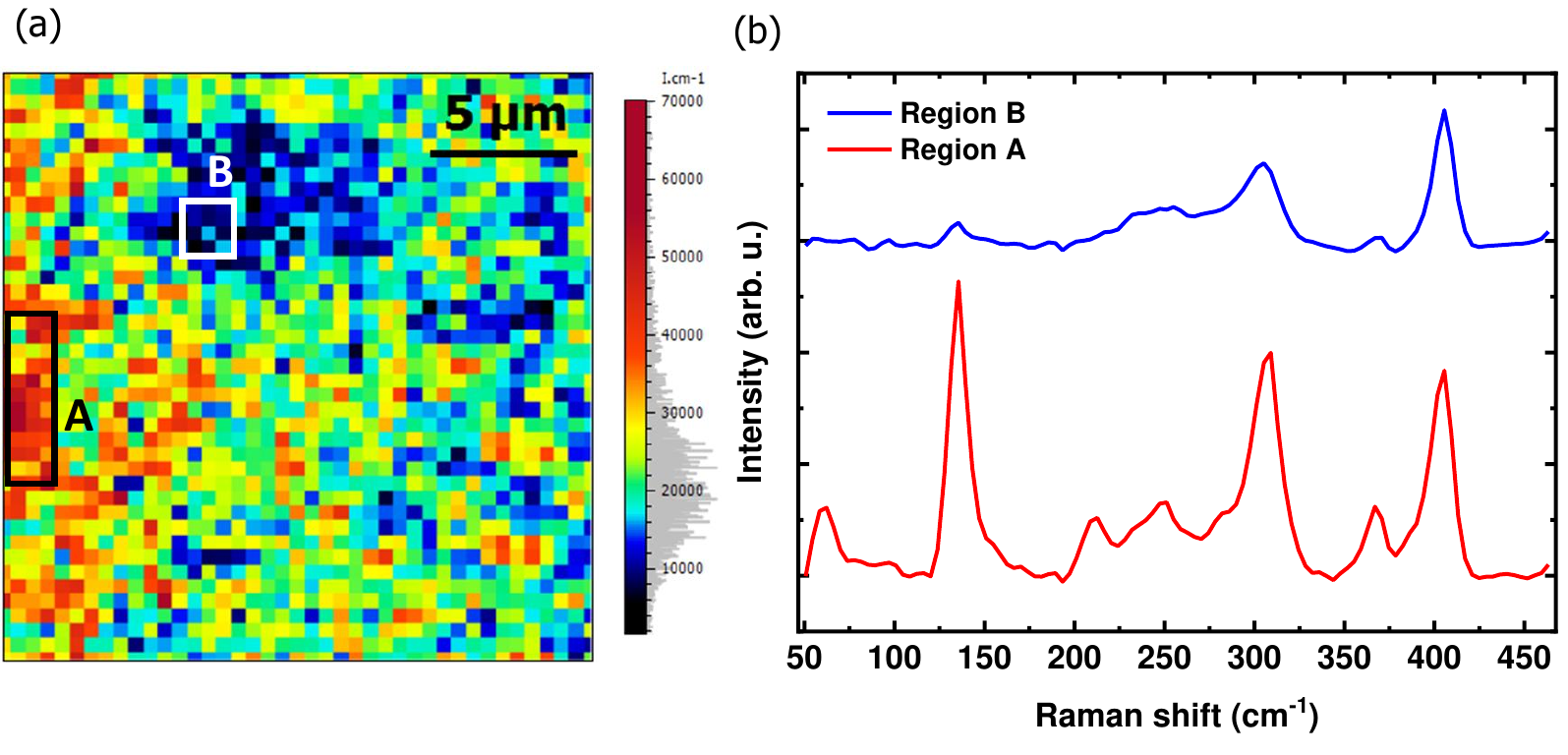}
\caption{\textbf{Spatially resolved Raman analysis of the 3-minute growth sample.} 
\textbf{a)} Display of the integrated intensity of the A$^{1}_{1g}$ mode of GaSe. The curve was fitted with a Lorentzian lineshape and a 5th-order polynomial background.
\textbf{b)}
 Mean Raman spectra extracted from the regions highlighted in a). Region~A corresponds to large integrated peak areas (red curve), while region~B corresponds to a weak signal intensity (blue curve).}
\label{fig1B}
\end{figure}

Raman mappings at room temperature were performed on both samples using a low-groove-density grating to enable simultaneous acquisition of the photoluminescence signal. Prior to mapping, a calibration study was conducted to determine the maximum laser power density that avoids local heating or material transformation. A threshold of $10^{6}_{}\,$W/cm$^2_{}$ was established. 

Due to the different growth durations, the 30-minute sample exhibits a significantly greater thickness (on the order of a few micrometers) compared to the 3-minute sample (a few nanometers), resulting in substantially higher signal intensities. This difference is particularly evident when comparing PL intensities and Raman signals. For the 3-minute sample, a measurement time of 1~second at a laser power density of $4\times10^{5}_{}\,$W/cm$^2_{}$ did not saturate the detector. In contrast, the detector saturated for the 30-minute sample already at an acquisition time of 0.0025~s. The Raman signal is likewise considerably stronger in the thicker sample.

Figure~\ref{fig1B}(a) shows the integrated area of the A$^{1}_{1g}$ mode of GaSe over a 20x20~\textmu m area with a stepsize of $500$~nm. The map reveals regions of high crystallinity as well as areas with negligible GaSe coverage. Representative mean spectra from these regions are presented in Figure~\ref{fig1B}(b). The spectrum corresponding to the high-crystalline region exhibits peaks at 58, 135, 209, 249, and 309~cm$^{-1}_{}$ which are characteristic of GaSe~\cite{Avdienko.2019, Diep.2019}. Owing to the lower spectral cutoff of the Raman-imaging instrument, the E$^{2}_{}$-low mode at 60~cm$^{-1}_{}$ is detectable. However, due to limited spectral resolution, it remains unclear whether the mode at 211~cm$^{-1}_{}$ is of E$^{2}_{1g}$ or E$^{2}_{2g}$ symmetry or a superposition of both. 

Additional bands at 368 and 405~cm$^{-1}_{}$ originate from the GaP substrate~\cite{langlotz2025mocvd}. The spectrum from the low-density region also shows GaSe-related bands at 135 and at 306~cm$^{-1}_{}$, indicative of GaSe. While the 135~cm$^{-1}_{}$ band is weak as expected, the 306~cm$^{-1}_{}$ band is more pronounced, likely due to the contribution of acoustic phonons from the Si/GaP substrate~\cite{langlotz2025mocvd}. The spectral region between 200-300~cm$^{-1}_{}$ collapses into a broad band around 250~cm$^{-1}_{}$ which may indicate a dominant contribution of the out-of-plane GaSe mode~\cite{Avdienko.2019, Rodriguez.2014}. This observation is consistent with a low GaSe density, potentially arising from sparse nucleation sites or enhanced contributions from the edges of adjacent two-dimensional GaSe nanosheets. No signatures of other polymorphs or phases of GaSe were detected in the 3-minute sample. 

These spatial variations in the Raman response motivate a closer examination of the corresponding optical emission behavior using PL spectroscopy. The room-temperature PL of the 3-minute sample is spatially inhomogeneous with 1-3 individual bands between 1.75-2.05~eV. The low-density region discussed above exhibits almost no PL, whereas the high crystallinity region shows a distinct signal composed of two bands between 1.8 and 2.0~eV. Regions of strong PL intensity spatially correlate with areas of high GaSe Raman signal. Corresponding PL maps and representative spectra are provided in Figure-Supp-Info-MicroPL3min. 

\begin{figure}
 \centering\includegraphics[width=12cm]{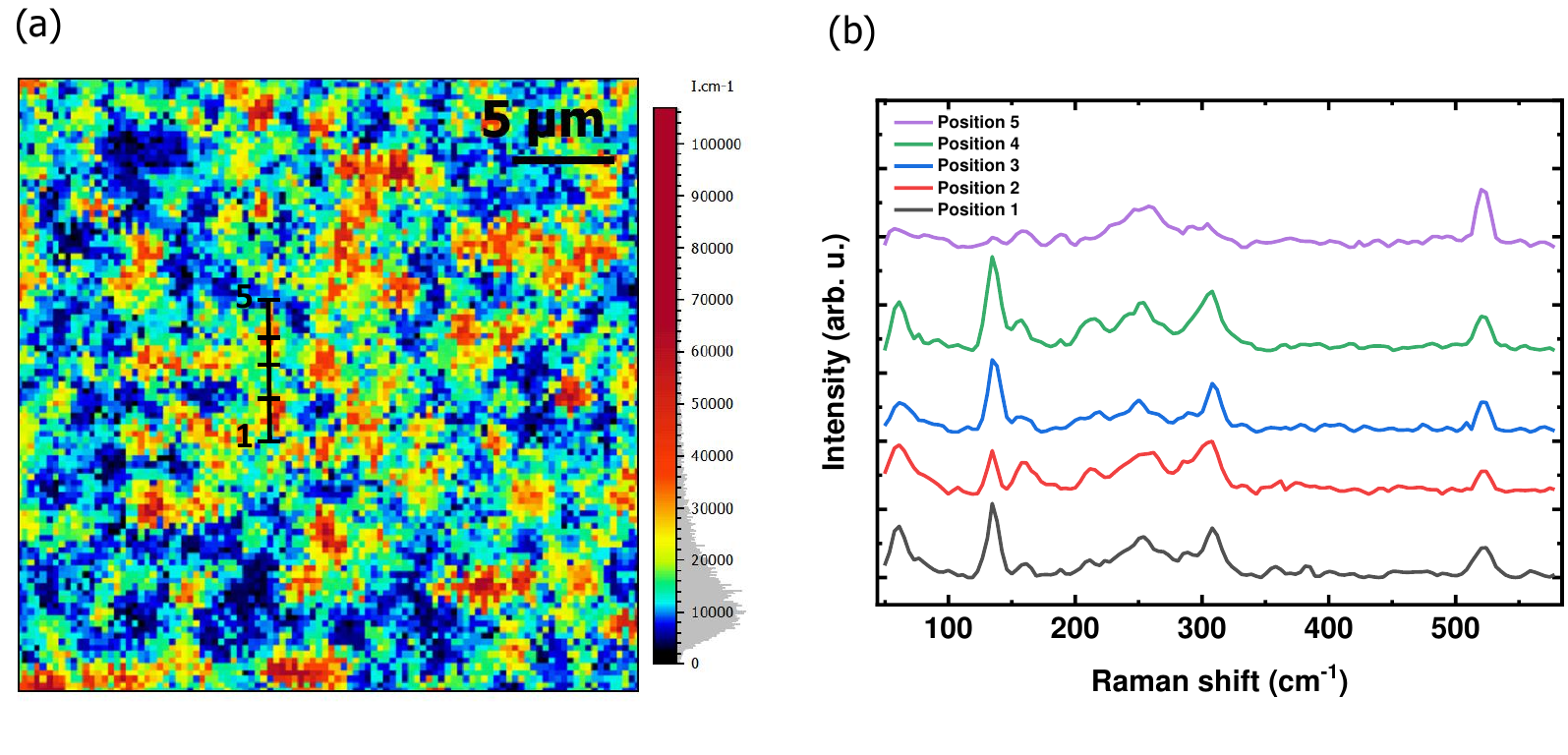}
\caption{\textbf{Spatially resolved Raman analysis of the 30-minute grown sample.} 
\textbf{a)} Display of the integrated intensity of the 157~cm$^{-1}_{}$ mode of Ga$_2$Se$_3$. The curve was fitted with a Lorentzian lineshape and a 5th-order polynomial background. The map has been processed with a median denoising filter of 25~\%.
\textbf{b)}
 Raman spectra extracted from the linescan highlighted in a). Spectra were chosen such that they alternate between high-intensity and low-intensity regions of the 157~cm$^{-1}_{}$ mode.}
\label{fig1C}
\end{figure}

The 30-minute sample exhibits fewer low-density regions, consistent with AFM and SEM observations, see Figure~\ref{fig1}. The lateral growth presumably covers the blank regions between the initially isolated GaSe islands. While most of the sample surface shows signatures of GaSe with the Raman bands assigned in Table~\ref{tab:raman_peaks}, additional bands occur at 157~cm$^{-1}_{}$ and around 290~cm$^{-1}_{}$ as a low-frequency shoulder of the 308~cm$^{-1}_{}$ GaSe-band. A map of the integrated intensity of the 157~cm$^{-1}_{}$ band is depicted in Figure~\ref{fig1C}. Compared to the 3-minute sample, the Raman signal exhibits stronger spatial variations. Therefore, a reduced step size of $300$~nm was used to better resolve these features. A linescan across high and low-intensity regions of the 157~cm$^{-1}_{}$ is shown in Figure~\ref{fig1C}. The additional bands are commonly assigned to Ga$_2$Se$_3$~\cite{SMIRI2024112256}. Their occurrence in the vicinity of regions with high GaSe E$^{2}_{1g}$ signal indicates the presence of local defects formed during coalescence of initially phase-pure GaSe islands, or post-growth aging effects during the oxidation process, see also Figure~SI-Raman. Reportedly weaker modes of Ga$_2$Se$_3$ at 105 and 118 cm$^{-1}_{}$~\cite{langlotz2025mocvd} have not been detected. Although thermal decomposition was minimized by carefully controlling the laser power density, it can not be completely excluded.

\begin{figure}[h]
 \centering\includegraphics[width=0.9\textwidth]{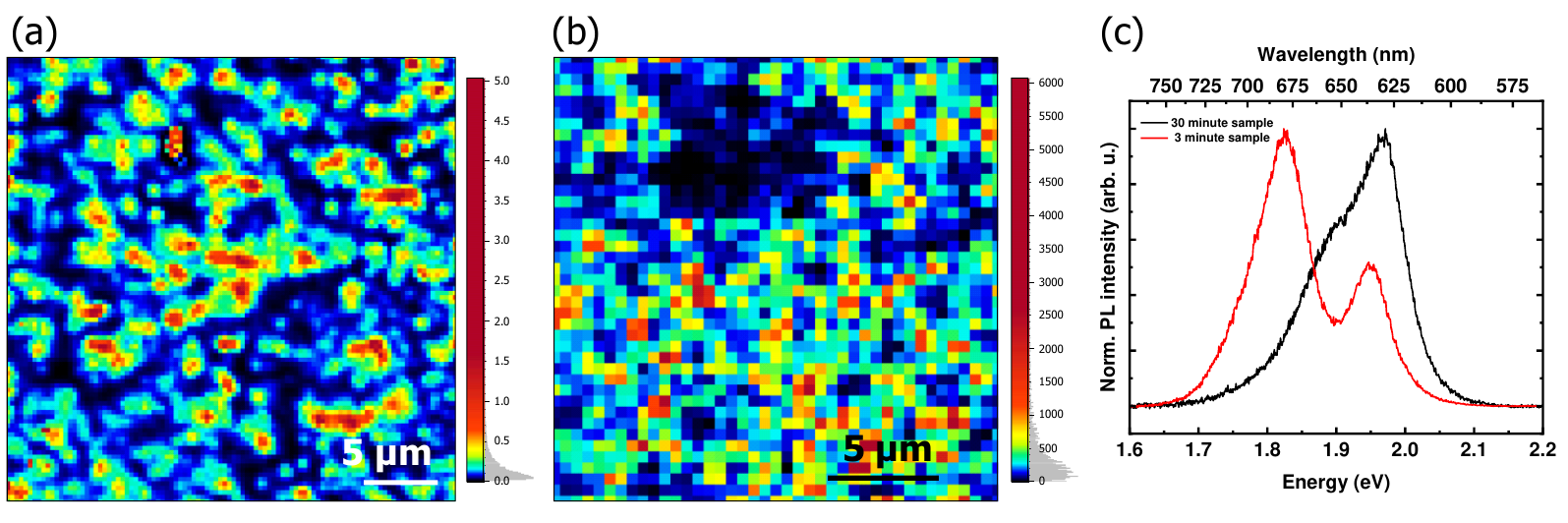}
 \caption{\textbf{Room-temperature PL characterization of GaSe samples:} \textbf{a)} PL map showing the integrated area of the PL bands of the 3 minute grown GaSe sample. \textbf{b)}  PL map showing the integrated area of the PL bands of the 30 minute grown GaSe sample. \textbf{c)} Average PL spectra of the two samples.}
\label{fig_Raman_PL}
\end{figure}

The room-temperature photoluminescence of the 30-minute sample occurs in the same energy range as that of the 3-minute sample, i.e., between 1.75-2.05~eV.  However, the lower-energy emission band is strongly suppressed, and the spectrum is dominated by two higher-energy bands. The average PL spectrum of the 20x20~\textmu m maps is shown in Figure~\ref{fig_Raman_PL}. The dominant emission band in the 30-minute sample is the high-energy band around 1.95~eV, whereas the 3-minute sample emission is most prominent around 1.85~eV. 

An intercorrelation analysis between Raman and PL maps (Figure SI-Correlate) reveals that, for the 3-minute sample, the GaSe Raman band at 135~cm$^{-1}_{}$ is weakly correlated with the PL intensity. While large-scale spatial features overlap, localized PL maxima do not necessarily coincide with Raman intensity maxima. This is particularly evident when comparing regions A and B in Figure~\ref{fig1B}(a) to ~Figure~\ref{fig_Raman_PL}(a). 

In contrast, the 30-minute sample shows a clearer correlation between PL emission and the GaSe band at 135~cm$^{-1}_{}$ and the Ga$_2$Se$_3$ band at 157~cm$^{-1}_{}$. However, as indicated in Figure~SI-Correlate b) and c) the PL emission predominantly originates from the GaSe regions rather than the Ga$_2$Se$_3$-rich areas. To reflect the local inhomogeneity of the emission, a line scan in comparison to Figure~\ref{fig1C} was extracted and is depicted in Figure~7 of the SI. Even though the Raman line scan highlights two spots with Ga$_2$Se$_3$ signatures, the photoluminescence signal splits in two bands only for one of these spots. This finding suggests that the Ga$_2$Se$_3$-rich regions in this sample are of different origins.

\subsection{Spatially resolved cathodoluminescence emission}

\begin{figure}[h]
 \centering\includegraphics[width=0.7\textwidth]{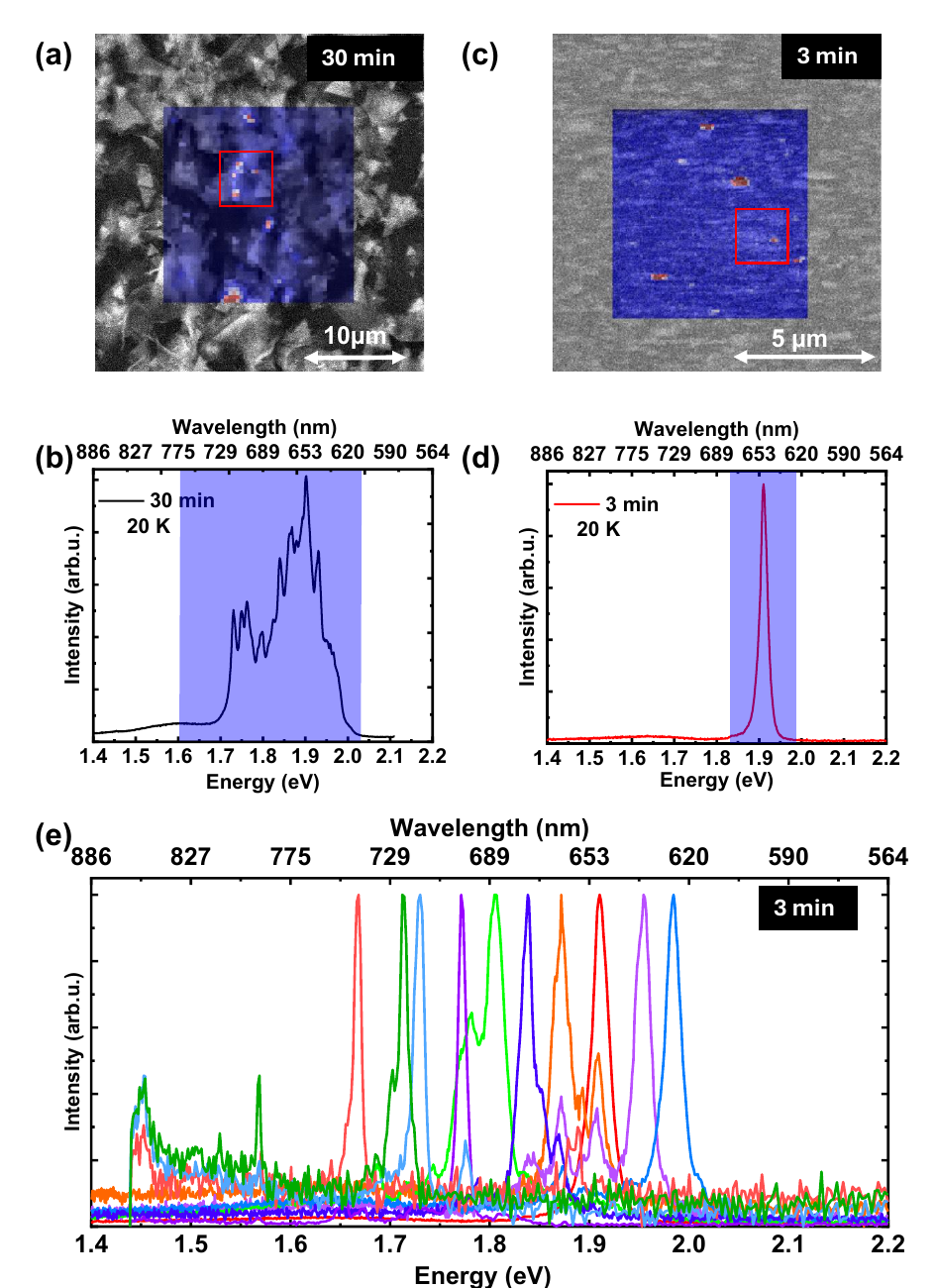}
 \caption{\textbf{CL characterization of GaSe samples:} \textbf{a)} CL intensity map of the 30-minute-grown GaSe sample at low temperature (20 K) with an exposure time of 100 ms. The red square indicates the region from which the spectrum was extracted. \textbf{b)}  Extracted CL spectrum at 20 K from the selected area in (a), showing the emission peak of the 30 min-grown sample. \textbf{c)} CL intensity map of the GaSe sample grown for 3 min, with the red square marking the region used for spectral extraction.  \textbf{d)} Corresponding extracted CL spectrum at 20 K from the selected region in (c), displaying the emission peak of the 3 min-grown sample.  \textbf{e)} Overlaid CL spectra collected at different spots on the 3 min-grown GaSe sample, highlighting the spatial variation of emission peaks.}
\label{fig2}
\end{figure}

To gain further insight into the spatial origin of the emission, low-temperature CL mapping was employed. CL mapping was performed for both 30-minute and 3-minute-grown GaSe sample at 20~K, enabling the simultaneous acquisition of spatially resolved emission and structural information via SEM imaging, as depicted in Figure~\ref{fig2}(a),(b).
For the 30-minute sample, the CL intensity map (Figure \ref{fig2}(a)) reveals spatially localized emission spots across the scanned region, with characteristic spot sizes on the order of $\sim$300–500 nm. The corresponding spectrum (Figure \ref{fig2}(b)) is dominated by a broad emission band extending from approximately 620 to 730 nm (1.7–2.0 eV). Weakly resolved excitonic sharp features with spectral widths ranging between 5 and 15 meV are superimposed on this broad background, indicating contributions from multiple recombination channels. This broadband response is consistent with increased structural disorder, enhanced defect-mediated recombination pathways, and reduced excitonic confinement in thicker GaSe layers~\cite{dwedari2019disorder, langlotz2025mocvd}. 
In contrast, the 3-minute sample exhibits pronounced spatial and spectral localization of the emission. The CL map (Figure \ref{fig2}(c)) shows isolated and bright spots with typical lateral dimensions of $\sim$300 nm and an estimated emitter density of $(8 \times 10^{7})\,\mathrm{cm}^{-2}$. A representative localized emitter (red square in Figure \ref{fig2}(c)) was selected for spectral analysis. The corresponding spectrum (Figure \ref{fig2}(d)) displays a spectrally narrow and intense emission peak centered at 1.91 eV (649 nm), with a FWHM of approximately 20 meV. Such linewidths are characteristic of strongly localized excitonic recombination and can be comparable to defect-bound exciton emission reported in other layered materials \cite{wu2026probing,shima2024cathodoluminescence}. Although spatial localization of emission is observed in both samples, their spectral characteristics differ significantly. This suggests that increasing the film thickness from a few nanometers to the micrometer regime substantially enhances surface roughness and promotes disorder-induced spectral broadening, while thinner layers favor the formation of discrete, localized excitonic states. To further assess spectral inhomogeneity within the 3-minute sample, CL spectra were extracted from multiple positions within the mapped region. As shown in Figure \ref{fig2}(e), several discrete emission peaks appear in the energy range from 1.65 to 1.99 eV. The observed variation in emission energy and number of peaks reflects local fluctuations in the electronic potential landscape, which may arise from strain variations, thickness fluctuations, or defect-induced confinement potentials~\cite{Budweg2019, sun2025designable, varghese2025quantum, varghese2025strain}. Overall, these results corroborate that the excitonic emission landscape in MOCVD-grown GaSe is strongly governed by growth duration, which controls both the structural uniformity and the degree of exciton localization. The emergence of spatially and spectrally isolated emitters in the thinner sample highlights the potential of GaSe for localized light emission, providing a foundation for the (quantum) optical investigations discussed in the following section.

\subsection{Optical emission properties and single-photon emission from localized states}

Further insight into the optical properties and the nature of the localized emission is obtained by $\mu$PL measurements were conducted on both samples. The detailed experimental configuration is provided in the Methods Section 4.3 and the SI Section 4.

\begin{figure}[!h]
 \centering\includegraphics[width=0.9\textwidth]{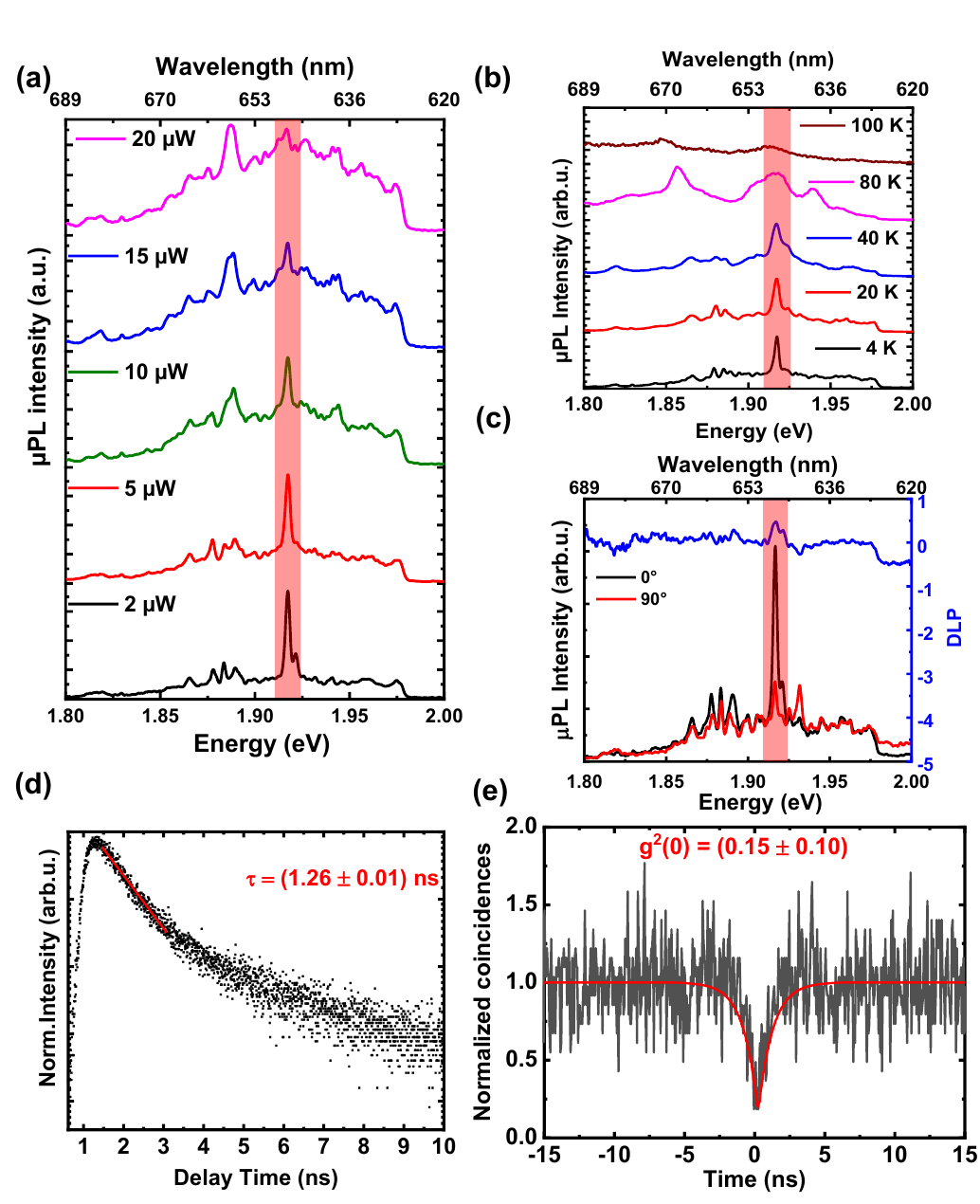}
 \caption{\textbf{Optical characterization of the 3 minute grown GaSe sample. } \textbf{a)} Power-dependent $\mu$PL spectra of the 3 min-grown GaSe sample measured at low temperature, with excitation power varying from 2 to 20 $\mu$W, showing the evolution of the emission peak intensity and spectral features. \textbf{b)} Temperature-dependent $\mu$PL spectra recorded from 4 K to 100 K, illustrating the thermal evolution of the emission peak, including changes in intensity and linewidth. \textbf{c)}  Polarization-resolved $\mu$PL spectra measured at 4 K for detection angles of 0$\degree$ and 90$\degree$, together with the corresponding DLP plotted on the right axis. \textbf{d)} TRPL decay curve measured at 4 K, fitted with a single-exponential function yielding a lifetime of $\tau$ = (1.26 $\pm$ 0.01) ns. \textbf{e)} Second-order autocorrelation function $g^{(2)}(0)$ = (0.15 $\pm$ 0.10), measured at 4 K under CW excitation.  }
\label{fig3}
\end{figure}

Excitation power–dependent $\mu$PL measurements of the 3-minute sample reveal a systematic redshift of the emission peak as the excitation power increases from 2~$\mu$W to 20~$\mu$W as shown in Figure \ref{fig3}(a). A comparable trend is also observed at higher temperatures like 20~K, 40~K, and 80~K (see SI, Figure 3(a)). In systems exhibiting localized emission, such excitation-dependent redshifts are commonly attributed to carrier-induced screening of localized confinement potential and bandgap renormalization~\cite{dai2025giant, sun20172d}. These effects collectively lead to a reduction in the emission energy with increasing carrier density. Similarly, the FWHM of the emission peak increases with excitation power. At 4~K, the linewidth broadens by nearly a factor of three, from ($2.7 \pm 0.8$)~meV to ($7.8 \pm 5.1$)~meV, with increasing excitation power, while similar broadening behavior is observed at elevated temperatures (see SI, Figure 3(b)). This broadening effect reflects enhanced carrier–carrier scattering and stronger many-body interactions within the localized states at higher excitation densities~\cite{dai2025giant}. Moreover, the integrated emission exhibits a sub-linear dependence of 0.49 $\pm$ 0.01 with excitation power, consistent with progressive saturation of the localized emitting centers or defect-mediated recombination~\cite{langlotz2025mocvd}.

When considering the dependence of excitation-power as a function of temperature, the $\mu$PL intensity follows a power-law behavior, I\textsubscript{PL} $\propto$ P\textsuperscript{$\alpha$}, with the exponent $\alpha$ increasing systematically with temperature. At 4~K, the strongly sublinear behavior indicates efficient carrier trapping in defect-induced potential minima in multilayer GaSe. As the temperature increases, $\alpha$ approaches unity as shown in SI Figure 3(d), suggesting thermal activation of excitons out of localized states and a reduced influence of trap-limited recombination~\cite{wei2016bound, bietti2012controlled, yuan2018interplay}. This evolution is accompanied by a temperature-induced redshift of the emission energy and spectral broadening, as illustrated in Figure~\ref{fig3}(b) and SI, Figure 3. These effects arise from bandgap renormalization and enhanced exciton–phonon interactions, phenomena that are well established in other 2D material systems~\cite{moody2016exciton, anagha2025unveiling, sahoo2026robust}. Despite the increase in $\alpha$, the integrated $\mu$PL intensity decreases with temperature due to the thermal activation of nonradiative recombination pathways. Together, these observations consistently indicate that the emission originates from strongly localized excitonic states, likely associated with defect-induced potential minima in the GaSe multilayer.

To further probe the nature of the localized emission, polarization-resolved $\mu$PL measurements were performed at 4~K, as shown in Figure \ref{fig3}(c). The emission peak exhibits an energy shift of approximately 170~$\mu$eV between orthogonal detection angles, accompanied by a markedly higher intensity at 0$\degree$ polarization. The extracted degree of linear polarization (DLP) for this emission is approximately 0.5. The observed polarization-dependent intensity, together with the small energy splitting and the elevated DLP, indicates an anisotropic optical transition associated with a localized emitting state~\cite{suzuki2018effect, wang2021highly}.
Similar excitation power-dependent and polarization-resolved $\mu$PL measurements were also performed for the 30-minute-grown sample at 4~K, and the corresponding results are presented in SI, section~3.

Beyond the spectral properties, TRPL measurements provide further insight into the recombination dynamics of the localized emitters. Figure~\ref{fig3}(d) shows the decay trace of the selected emitter, which exhibits a biexponential behavior. The dominant decay component yields an excited-state lifetime of $(1.26 \pm 0.01)~\mathrm{ns}$. This value exceeds the lifetimes reported for GaSe grown by MBE ($\sim 400$ ps)~\cite{Rakhlin2024}, where faster recombination has been attributed to stronger nonradiative channels or carrier reduced confinement. At the same time, the measured lifetime remains shorter than those reported for exfoliated or strain-engineered GaSe systems~\cite{Tonndorf2017, Luo2023}, where lifetimes of several nanoseconds have been observed due to stronger confinement and reduced dielectric screening. The intermediate lifetime observed here therefore suggests a distinct confinement regime governed by defect-induced localization, providing sufficient spatial confinement while maintaining efficient radiative recombination.

To assess the quantum-optical nature of the emission from these spatially localized centers in the 3-minute MOCVD-grown sample, second-order autocorrelation measurements were conducted using an HBT setup. Figure~\ref{fig3}(e) shows the normalized second-order correlation function $g^{(2)}(\tau)$. A pronounced antibunching dip is observed at zero time delay ($\tau = 0$) for an excitation power of 2~$\mu\mathrm{W}$, indicating a strong suppression of multi-photon events and providing clear evidence of non-classical light emission. A fit to the experimental data (red curve) yields a value of
\[
g^{(2)}(0) = (0.15 \pm 0.10),
\]
which lies well below the threshold of 0.5 for single-photon emission. Compared to previously reported GaSe-based single-photon emitters in exfoliated and strain-engineered systems, which exhibit $g^{(2)}(0)$ values of $0.33$~ \cite{Tonndorf2017}, $0.43$~\cite{Luo2023b}, and $0.32$~\cite{Luo2023}, the emitter investigated here exhibits a substantially stronger antibunching response. The improved single-photon purity may arise from enhanced spatial isolation of the emitting center, reduced background fluorescence, or a lower probability of multi-exciton generation under non-resonant excitation. Collectively, these results indicate that the localized states formed during the short MOCVD-growth process act as efficient quantum confinement centers with minimal coupling to nearby emitters. However, the measured value remains slightly above the pulsed $g^{(2)}(0) = 0.12$ reported for Ga$_2$Se$_3$/GaSe nanostructures grown by MBE via van der Waals epitaxy~\cite{Rakhlin2024}. Noteworthy, in that work, the emitters were embedded in a carefully engineered heterostructure environment, which effectively suppresses surface trap states and minimizes environmental charge fluctuations. In contrast, the uncapped GaSe emitters investigated here are more susceptible to surface-related charge noise, adsorbates, and trap-assisted recombination pathways, which can induce spectral diffusion, blinking, or re-excitation processes and consequently a modest increase in the measured $g^{(2)}(0)$ value. 

These considerations indicate that the single-photon purity demonstrated here is not a fundamental limitation of the MOCVD-grown GaSe platform, but can likely be further improved through dielectric encapsulation or surface passivation strategies. In particular, encapsulation with hexagonal boron nitride or integration into engineered photonic environments may suppress spectral diffusion, enhance emission stability, and improve photon extraction efficiency. Beyond these improvements, the demonstrated wafer-scale growth and the emergence of spatially localized emitters highlight the potential of GaSe as a scalable platform for quantum photonics. Future efforts may focus on deterministic control of emitter formation, for example through strain engineering or controlled defect introduction, as well as integration with photonic cavities or waveguides to enable on-chip single-photon sources and to explore stimulated emission and lasing phenomena.

\section{Conclusion}\label{sec3}

In this work, we demonstrated that wafer-scale MOCVD-grown 2D GaSe on GaP/Si supports spatially localized excitonic emission with clear signatures of non-classical light generation. Structural characterization confirms phase-pure, epitaxially aligned GaSe, while optical measurements reveal a strong dependence of emission behavior on growth duration and the resulting multilayer morphology. In particular, short-growth GaSe layers exhibit pronounced spatial and spectral localization, excitation-dependent bandgap renormalization, anisotropic polarization, and nanosecond exciton lifetimes, consistent with defect-induced quantum confinement. Most importantly, second-order correlation measurements confirm single-photon emission with strong antibunching from uncapped MOCVD-grown GaSe, demonstrating that high single-photon purity can be achieved without encapsulation or strain engineering. The observed performance positions MOCVD-grown GaSe between exfoliated systems and fully engineered heterostructures, highlighting the role of growth-controlled defect landscapes in tailoring excitonic confinement. These results establish wafer-scale 2D GaSe as a promising platform for integrated classical and quantum light sources compatible with silicon-based architectures. Further optimization through surface passivation or dielectric encapsulation is expected to enhance spectral stability and photon purity. In combination with deterministic emitter control and integration into photonic cavities, this approach provides a pathway toward scalable layered-semiconductor quantum photonics and the exploration of advanced light–matter interactions, including cavity-enhanced emission and lasing phenomena.

\section{Methods}\label{sec4}
\subsection{MOCVD Growth of GaSe}
GaSe layers were grown by MOCVD in an AIXTRON AIX 200 GFR horizontal reactor, a schematic illustration of which is shown in Figure S1. This well-established reactor design has proven useful for the reliable synthesis of complex material systems, such as (GaIn)(NAs)~\cite{VOLZ20092418}, Ga(NAsP)~\cite{Kunert2006}, Ga(PAsBi)~\cite{NATTERMANN2016209} or (Ga,In)(As,Bi)~\cite{10.1063/1.5097138}. (100)-oriented GaP/Si templates~\cite{GaP_devolpment_Si} were used as substrates. Prior to GaSe deposition, a $4\,$nm GaP buffer layer was deposited at $675\degree$C using triethylgallium (TEGa) and tert-butylphosphine (TBP) to obtain a smooth, stabilized surface. After buffer growth, the reactor temperature was reduced to $600\degree$C, and a short phosphorus supply step was introduced to stabilize the GaP surface before initiating GaSe growth. The gallium and selenium precursors were TEGa and di-iso-propyl-selenide (DiPSe) (provided by Dockweiler Chemicals). Bubblers were maintained at $20\degree$C, with a total reactor pressure of $50\,$mbar and a gas flow of $6800\,$sccm. The TEGa partial pressure was kept constant at $3.5 \cdot 10^{-3}_{}\,$mbar, while the selenium supply was kept at $9.45 \cdot 10^{-3}_{}\,$mbar corresponding to a VI/III ratio of $2.7$ and phase-pure GaSe, with a growth time of 3 and 30 minutes. Growth was carried out in continuous mode with simultaneous Ga and Se supply. Additional information on the reactor setup is provided in the SI (Section 1).

\subsection{Structural characterization}
The structural properties of the MOCVD-grown GaSe layers were investigated using a combination of complementary microscopy and spectroscopy techniques to probe morphology, crystallinity, and epitaxial quality across multiple length scales. Surface morphology and thickness were first examined by AFM operated in tapping mode, providing high-resolution topographical images and height profiles of the GaSe nanosheets. These measurements were complemented by SEM, which enabled the analysis of lateral morphology and growth orientation over larger areas. AFM measurements were performed using an Alphacen 300 system (Nanosurf), with the raw data processed and analyzed using Gwyddion software. Commercially available silicon probes (MikroMasch Europe) with a nominal tip radius of < 7 nm were used. SEM imaging was carried out using a Thermo Fisher Scientific Helios 5 Hydra CX operated in Ultra-High Resolution (UHR) immersion mode, with an acceleration voltage of 4 kV and a beam current of 0.1 nA.

To assess the crystalline quality and phase purity of the layers, Raman spectroscopy was applied using a Horiba XPloRA Plus confocal Raman microscope equipped with a 100×/0.8 NA objective, resulting in a lateral resolution of approximately 1 $\mu$m. The presented spectra correspond to the average of a $21 \times 21$ point Raman map acquired over an area of $40 \times 40\,\mu$m$^{2}_{}$. A continuous-wave excitation laser ($\lambda$ = 532 nm) with a diffraction grating of 2400 grooves/mm, yielding a spectral resolution of $\pm 1\,$cm$^{-1}$. The applied laser power density was typically on the order of $10^{5}_{}\,$W/cm$^2_{}$. The observed vibrational modes were compared with reported GaSe signatures to verify phase formation and structural integrity. 

Further insight into the crystal structure and heterointerface was obtained by high-resolution TEM, including STEM, performed using a JEOL JEM-2200FS microscope operated at an acceleration voltage of 200 kV. These measurements allowed direct visualization of the stacking sequence and revealed the epitaxial relationship between the GaSe layer and the underlying GaP/Si substrate, providing evidence of atomic-scale structural coherence. Finally, XRD measurements were carried out using a Panalytical X’Pert Pro diffractometer equipped with Cu K$\alpha$ radiation ($\lambda = 1.5406,\mathrm{\AA}$) to confirm the phase purity and crystallographic orientation of the GaSe layers, as well as to extract lattice parameters, thereby complementing the local structural analysis with ensemble-averaged information. More details on the structural characterizations are provided in the SI (Section 2).

Spatially resolved Raman and photoluminescence spectroscopy was performed using a Thermo Fisher Scientific DXR3xi Raman-Imaging microscope equipped with an EMCCD camera, an x100 Olympus MPlan objective with NA = 0.90, and a 532~nm diode-pumped solid-state laser, operating with $4\times10^{5}_{}\,$W/cm$^2_{}$ to prevent local heating. A 400~lines/mm grating with a spectral resolution of 11~cm$^{-1}_{}$ was used to allow simultaneous collection of Raman and PL signals. The spatial resolution of the setup was better than 500~nm and controlled with a $25$~\textmu m aperture. Exposure times were set to $2\times1$~s for the Raman measurements and $8\times0.025$~s for the PL measurements of the 30~minute sample.

\subsection{Optical characterization}
The optical properties of the GaSe samples were investigated using a combination of CL and  (\textmu PL spectroscopy, providing both spatially resolved and spectrally resolved insight into the emission characteristics. CL measurements were performed at low temperature (20 K) in a scanning electron microscope equipped with a CL detection system. This configuration enables the simultaneous acquisition of SEM images and spatially resolved emission maps, allowing a direct correlation between structural features and optical response across the sample. Complementary \textmu PL measurements were carried out using a confocal setup integrated with a closed-cycle cryostat and a three-axis piezoelectric positioning stage. The samples were excited by a continuous-wave laser ($\lambda$ = 532~nm), focused to a spot size of approximately 3–5~\textmu m. The emitted PL was spectrally analyzed in a monochromator equipped with a dispersing grating (300 grooves/mm) and detected using a silicon CCD camera, yielding a spectral resolution of approximately 350 \textmu eV. Based on this setup, power- and temperature-dependent \textmu PL measurements were performed to investigate the emission mechanisms and excitonic behavior, with temperatures ranging from 4 to 100~K. In addition, TRPL measurements were employed to extract recombination dynamics and exciton lifetimes associated with the localized emitters. To further probe the quantum nature of the emission, second-order autocorrelation measurements were conducted using a HBT interferometer under continuous-wave excitation. The normalized second-order correlation function $g^{(2)}(\tau)$ was evaluated to assess the single-photon emission characteristics. More details on the experimental setup are provided in the Figure 5 of the SI.

\backmatter

\bmhead{Acknowledgements}
We also thank Dockweiler Chemicals GmbH, Marburg, for collaboration regarding precursor development and purification.
\section*{Declarations}
\begin{itemize}
\item Funding:
The authors gratefully acknowledge financial support from the Deutsche Forschungsgemeinschaft (DFG) through SFB 1083 (Project No. 223848855) and SPP 2244 (Project No. 443416027), as well as within the framework of the LOEWE 3 project “1for2D – Single Source Precursors for 2D Materials” (HA Project No. 2140/25–249). The authors further acknowledge funding from the European Regional Development Fund (ERDF) and the Recovery Assistance for Cohesion and the Territories of Europe (REACT-EU). Additional financial support from the Berlin Senate through Berlin Quantum (BQ) is also gratefully acknowledged.

\item Author contribution:
B. S. S. and N. F. L. contributed equally to this work. B. S. S. led the $\mu$PL characterization, HBT measurements, and data evaluation, with support from S. M., A. K.-S., C. C. P., and I. L. Spatially resolved room-temperature Raman and PL measurements were performed and evaluated by M. R., who also wrote the corresponding sections of the manuscript and Supporting Information. N. F. L., with the assistance of R. G., M. B., and N. G., carried out the epitaxial growth of the investigated samples by MOCVD, as well as the morphological characterization using AFM, SEM, XRD, and Raman spectroscopy. J. B. conducted and evaluated the TEM measurements. I. L. and K. G. performed the CL measurements. I. L. supervised the spectroscopy-related aspects of the project with support from S. R., while K. V. supervised the epitaxial growth and structural characterization. The manuscript was written by B. S. S. and N. F. L. with support from I. L., S. R., and K. V., and with input from all authors.

\item Data availability:
The data that support the findings of this study are available from the corresponding author upon reasonable request.

\item Materials availability:
Please refer to the supplementary material for additional details on the MOCVD reactor and optical setup. It also includes XRD measurements of the 3-minute-grown GaSe samples, as well as PL analysis, covering temperature- and power-dependent PL emission of the 3-minute sample and power-dependent PL measurements of the 30-minute-grown GaSe sample.

\end{itemize}

\bigskip


\bibliography{sn-bibliography}
\end{document}


\title[Article Title]{Supplementary Information: Localized Excitonic Emission in Wafer-Scale MOCVD-Grown GaSe 2D Nanosheets for Classical and Non-Classical Light Sources}

\author[1]{\fnm{Bhabani Sankar} \sur{Sahoo}}
\equalcont{These authors contributed equally to this work.}
\author[2,3]{\fnm{Nils Fritjof} \sur{Langlotz}}
\equalcont{These authors contributed equally to this work.}

\author[1]{\fnm{Shachi} \sur{Machchhar}}
\author[1]{\fnm{Kartik} \sur{Gaur}}
\author[2,3]{\fnm{Robin} \sur{Günkel}}
\author[2,3]{\fnm{Max} \sur{Bergmann}}
\author[2,3]{\fnm{Naghmeh} \sur{Ghadghooni}}
\author[1]{\fnm{Aris Koulas-} \sur{Simos}}
\author[2,3]{\fnm{Jürgen} \sur{Belz}}
\author[1]{\fnm{Chirag Chandrakant} \sur{Palekar}}
\author[1,4]{\fnm{Maximilian} \sur{Ries}}
\author[2,3]{\fnm{Kerstin} \sur{Volz}}
\author[1]{\fnm{Stephan} \sur{Reitzenstein}}
\author*[1]{\fnm{Imad} \sur{Limame}}\email{imad.limame@tu-berlin.de}

\affil[1]{\orgdiv{Institut für Physik und Astronomie}, \orgname{Technische Universit\"at Berlin}, \orgaddress{\street{Hardenbergstrasse 36}, \city{Berlin}, \postcode{10623}, \country{Germany}}}

\affil[2]{\orgdiv{mar.quest | Marburg Center for Quantum Materials and Sustainable Technologies}, \orgname{Philipps-Universität Marburg}, \orgaddress{\street{Hans Meerwein Str. 6}, \city{Marburg}, \postcode{35032}, \country{Germany}}}

\affil[3]{\orgdiv{Department of Physics, Structure and Technology Research Lab}, \orgname{Philipps-Universität Marburg}, \orgaddress{\street{Hans Meerwein Str.~6}, \city{Marburg}, \postcode{35032}, \country{Germany}}}

\affil[4]{\orgdiv{Chemical Analysis Division}, \orgname{Thermo Fisher Scientific GmbH}, \orgaddress{\street{Im Steingrund~4-6}, \city{Dreieich}, \postcode{63303}, \country{Germany}}}

\maketitle

\section{MOCVD reactor setup}
\begin{figure}
 \centering\includegraphics[width=9cm]{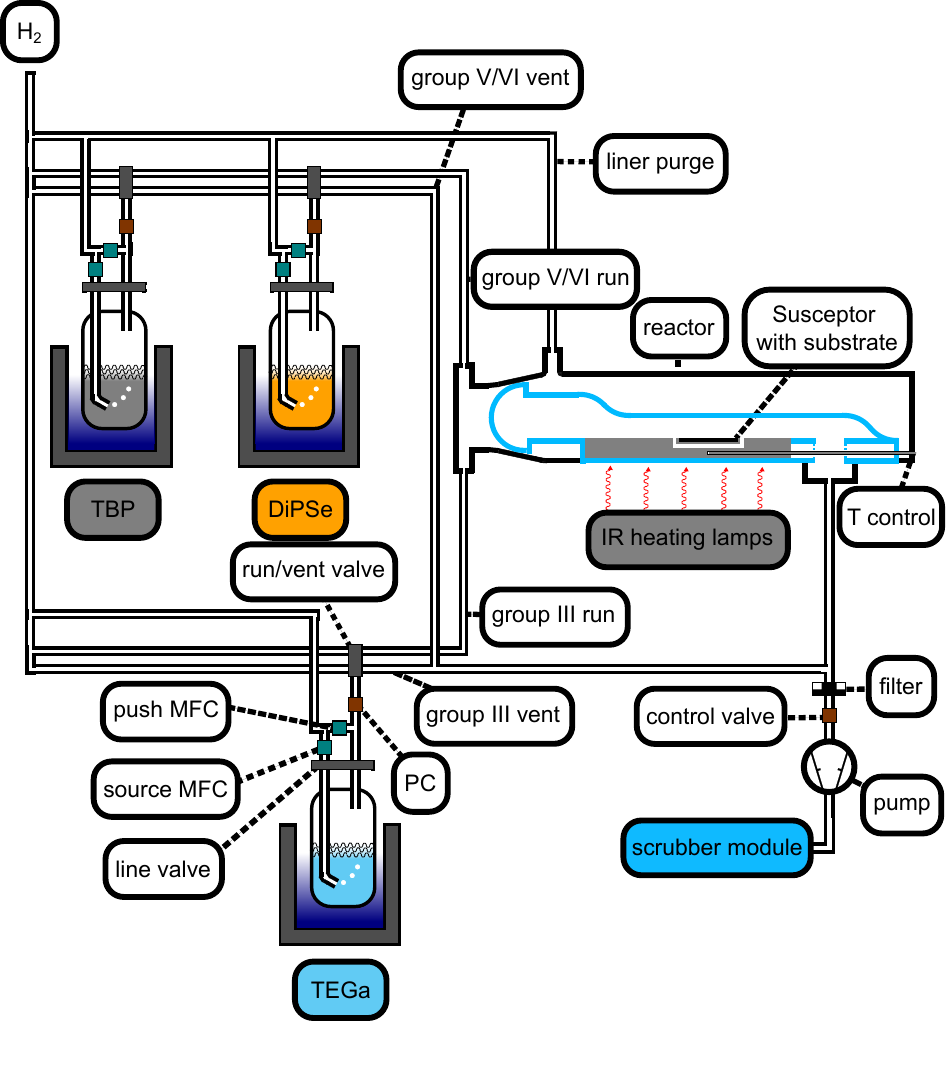}
 \caption{\textbf{MOCVD reactor setup:} Schematic showing the gas cabinet on the left and the reactor with exhaust and scrubber on the right.}
\label{figSX}
\end{figure}

Figure \ref{figSX} displays the MOCVD setup. Tertiarybutylphosphine (TBP), diisopropylselenide (DiPSe), and triethylgallium (TEGa) are used as group V, VI, and III precursors, respectively. Hydrogen (H$^{}_{2}$) flows through the bubblers, where it becomes saturated with precursor vapor and transports it into the reactor. Inside the reactor, the precursors decompose on the substrate as the temperature is increased by infrared heating lamps. Precursor species that are not incorporated into the growing layers are carried to the exhaust and removed by the scrubber. The precursor partial pressures are controlled by the bubbler bath temperature, the source mass flow controllers (MFCs), and the reactor pressure.

\section{X-ray diffraction of GaSe/GaP/Si}
\begin{figure}[!ht]
 \centering\includegraphics[width=0.9\textwidth]{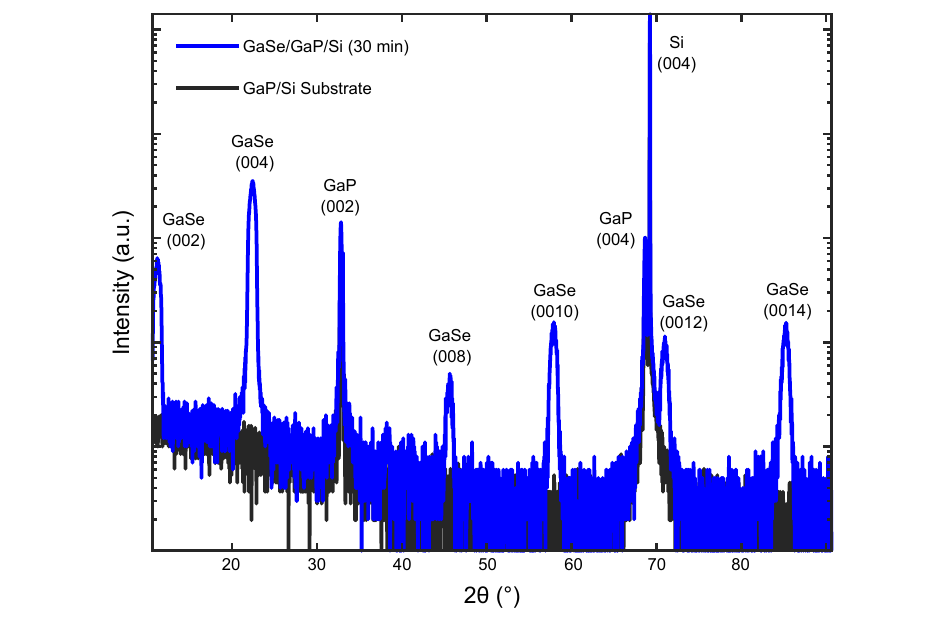}
 \caption{\textbf{X-ray diffraction (2$\theta-\omega$) patterns of GaSe/GaP/Si:} The violet spectrum corresponds to the GaP/Si substrate, showing the GaP (002) and (004) reflections together with the Si (004) peak. After 30 minute of GaSe growth, additional (00l) reflections of GaSe appear, indicating preferential c-axis oriented growth.}
\label{figxrd}
\end{figure}

The 2$\theta-\omega$ X-ray diffraction pattern shown in Figure \ref{figxrd} confirms the high structural quality of the heterostructure. The GaP/Si substrate (violet curve) exhibits the expected GaP (002) and (004) reflections together with the Si (004) peak, demonstrating the crystalline alignment of the underlying template. After 30 minute of GaSe growth, a series of GaSe (00X) reflections appears at 2$\theta$ of 11.3$^{\circ}_{}$, 22.42$^{\circ}_{}$, 45.76$^{\circ}_{}$, 57.93$^{\circ}_{}$, 72.06$^{\circ}_{}$, and 85.38$^{\circ}_{}$, which are indexed to the (002), (004), (008), (0010), (0012), and (0014) planes, respectively. The exclusive presence of (00X) reflections indicates strongly c-axis oriented growth of GaSe on GaP/Si. The XRD measurements were performed using a Panalytical X'Pert Pro diffractometer using Cu $K^{}_{\alpha}$ radiation.

From the peak positions, a lattice parameter of $c=15.89\pm0.10\,$\AA ~is determined. This value is in very good agreement with the reported literature value of $c=15.949\,$\AA ~\cite{Barker2025}, confirming phase-pure GaSe with negligible out-of-plane strain within the experimental uncertainty.

\section{Temperature-dependent power evaluation of PL of 30 minutes grown GaSe}

\begin{figure}[!ht]
 \centering\includegraphics[width=0.9\textwidth]{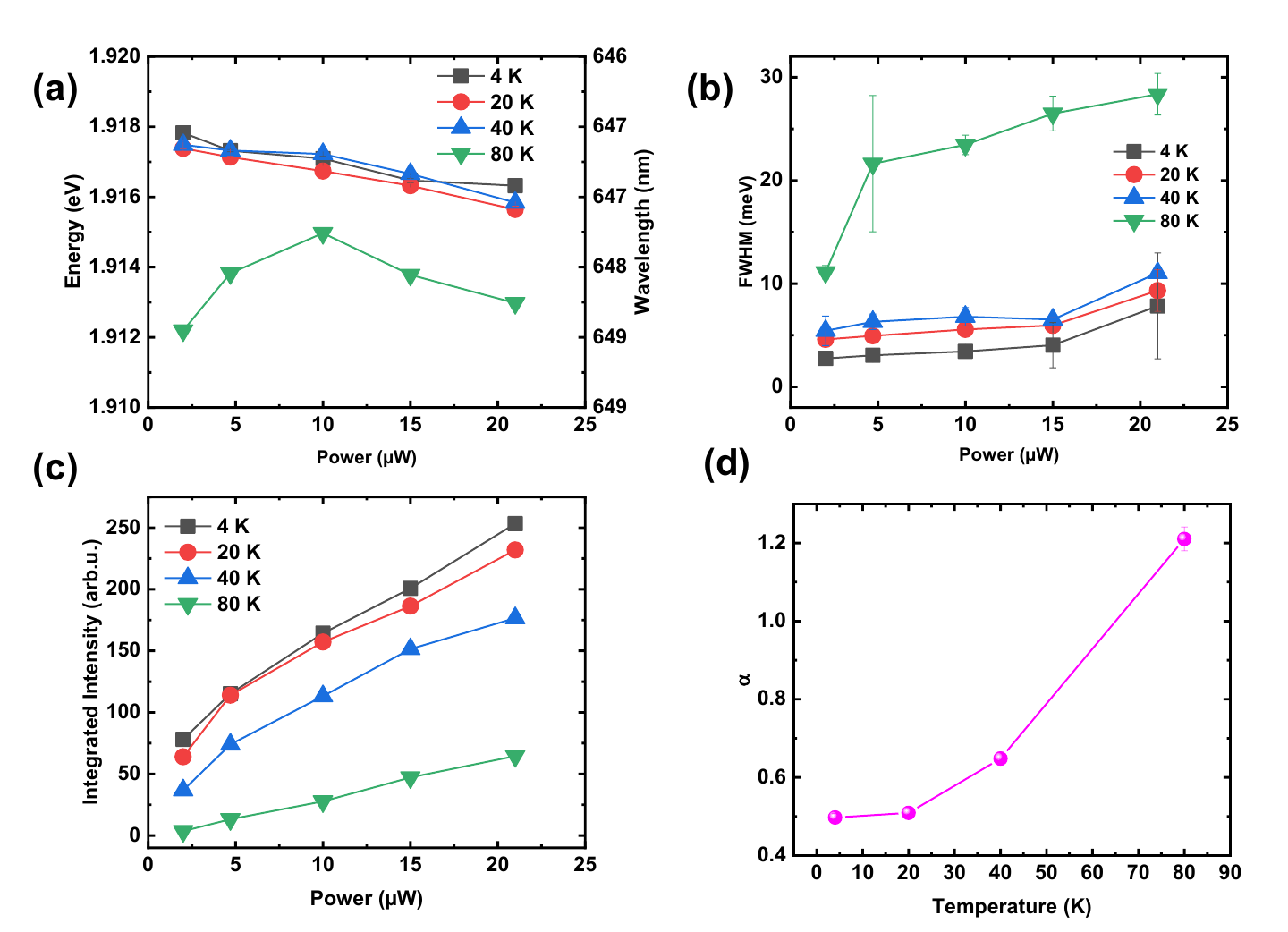}
 \caption{\textbf{Temperature dependent power evaluation of PL parameters for the 3 minutes grown GaSe sample} \textbf{a)}Extracted emission peak energy as a function of excitation power measured at 4 K, 20 K, 40 K, and 80 K. \textbf{b)} Corresponding FWHM variation with excitation power at different temperatures. \textbf{c)} Integrated PL intensity as a function of excitation power for the same temperature range. \textbf{c)} Temperature dependence of the power law coefficient.}
\label{fig4}
\end{figure}
\begin{figure}[h!t]
 \centering\includegraphics[width=0.9\textwidth]{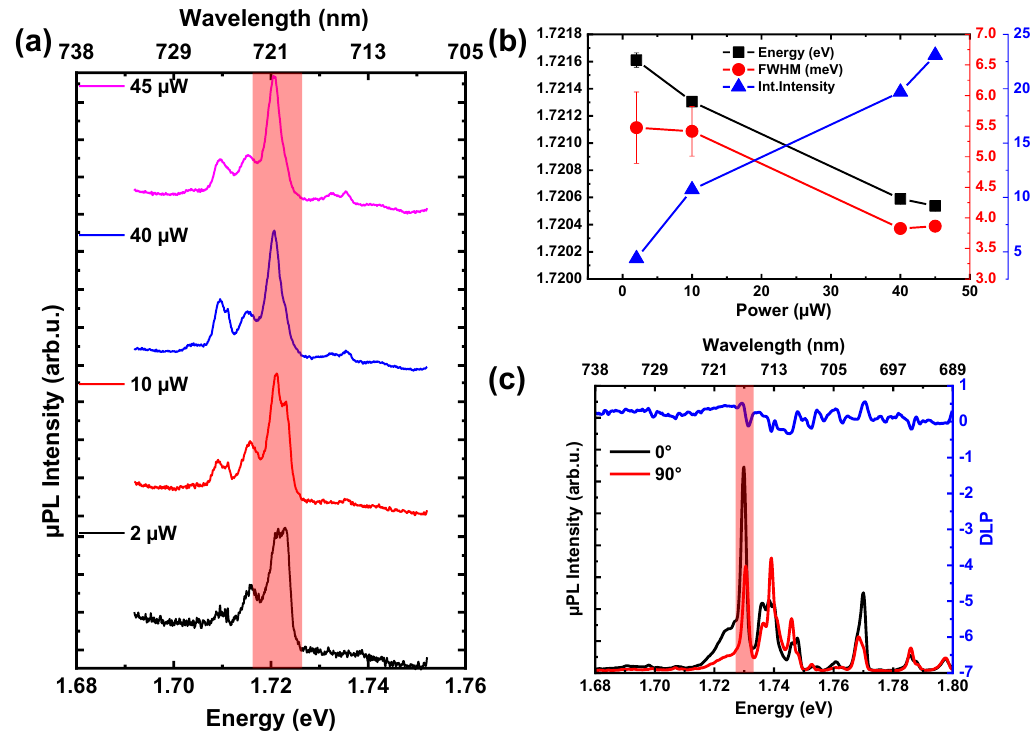}
 \caption{\textbf{PL characteristics of 30 minutes grown GaSe sample } \textbf{a)} Power-dependent PL spectra of the sample measured at excitation powers ranging from 2 to 45 µW  \textbf{b)} Extracted peak energy, FWHM, and integrated PL intensity as a function of excitation power \textbf{c)} Polarization-resolved $\mu$PL spectra acquired at 4 K for detection angles of 0$\degree$ and 90$\degree$, along with the corresponding DLP shown on the right axis.}
\label{fig5}
\end{figure}

In contrast to the 3-minute nanosheet sample, the 30-minute multilayer GaSe exhibits a micro-photoluminescence ($\mu$PL) emission peak that is redshifted relative to its corresponding CL emission, as shown in Figure \ref{fig5} (a). This energy offset is likely related to lateral spatial inhomogeneity within the thicker film, such as local variations in thickness, strain, or defect distribution. Since $\mu$PL and CL measurements are not necessarily performed at identical spatial locations and employ different excitation geometries, small bandgap fluctuations across the multilayer can give rise to measurable differences in observed emission energy. Similar to the 3-minute sample, the 30-minute sample also exhibits a systematic redshift with increasing excitation power at 4~K, which can be attributed to carrier-density–induced bandgap renormalization, as discussed in the main text. Figure \ref{fig5} (b) summarizes the extracted peak energy, FWHM, and integrated PL intensity as a function of excitation power, indicating a slight redshift of the emission energy and a gradual linewidth narrowing at higher powers, accompanied by a nearly linear increase in integrated intensity. In addition, polarization-resolved $\mu$PL measurements (Figure \ref{fig5} (c)) reveal that the emission detected at the 90$\degree$ polarization is blueshifted by approximately 610~$\mu$eV relative to the 0$\degree$ configuration, while the 0$\degree$ polarization exhibits a significantly higher emission intensity. Such polarization-dependent energy splitting and intensity contrast are indicative of an anisotropic optical transition, consistent with defect-related localized states in multilayer GaSe~\cite{suzuki2018effect, wang2021highly}.

\section{Micro-photoluminescence setup}
\begin{figure}[!h]
 \centering\includegraphics[width=0.8\textwidth]{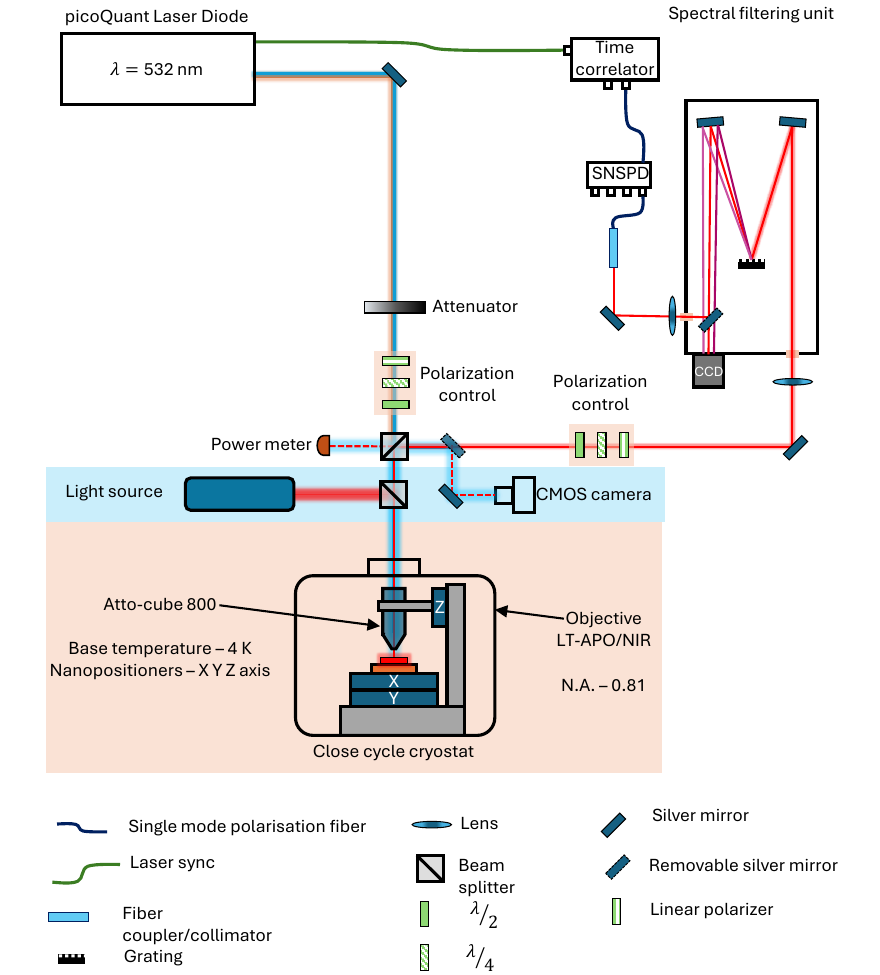}
 \caption{\textbf{Schematic of optical experimental setup } }
\label{fig6}
\end{figure}

The optical setup (Figure~\ref{fig6}) is based on a µPL configuration integrated with a closed-cycle cryostat. A 532 nm picoQuant laser diode is used as the excitation source, with the excitation power controlled by a variable attenuator and the polarization adjusted using a combination of half-wave ($\lambda$/2) and quarter-wave ($\lambda$/4) plates together with linear polarizers. The excitation beam is directed through a beam splitter and focused onto the sample mounted on XYZ nanopositioners inside an AttoCube 800 cryostat using a long-working-distance LT-APO/NIR objective (NA = 0.81). The emitted or reflected signal from the sample is collected by the same objective and routed either to a CMOS camera for real-space imaging or to a spectrally filtered detection arm. In the spectral analysis path, the collected emission is dispersed by a grating-based monochromator, where a charge-coupled device (CCD) camera records the wavelength-resolved photoluminescence spectra with high sensitivity and spectral resolution. For the time-resolved studies, removable silver mirrors and lenses guide the signal towards superconducting nanowire single-photon detectors (SNSPDs) connected to a time correlator. Single-mode polarization-maintaining fibers and fiber couplers/collimators are used throughout the detection system to ensure efficient signal collection and polarization control, while a power meter continuously monitors the excitation intensity. These are connected to a time correlator for photon-correlation measurements ($g^{(2)}(\tau)$) via single-mode polarization-maintaining fibers. A power meter and laser sync line ensure continuous monitoring of excitation stability and timing.

\newpage

\newpage
\section{Spatially-resolved Raman and PL measurements at room-temperature}
\begin{figure}[h!t]
 \centering\includegraphics[width=5cm]{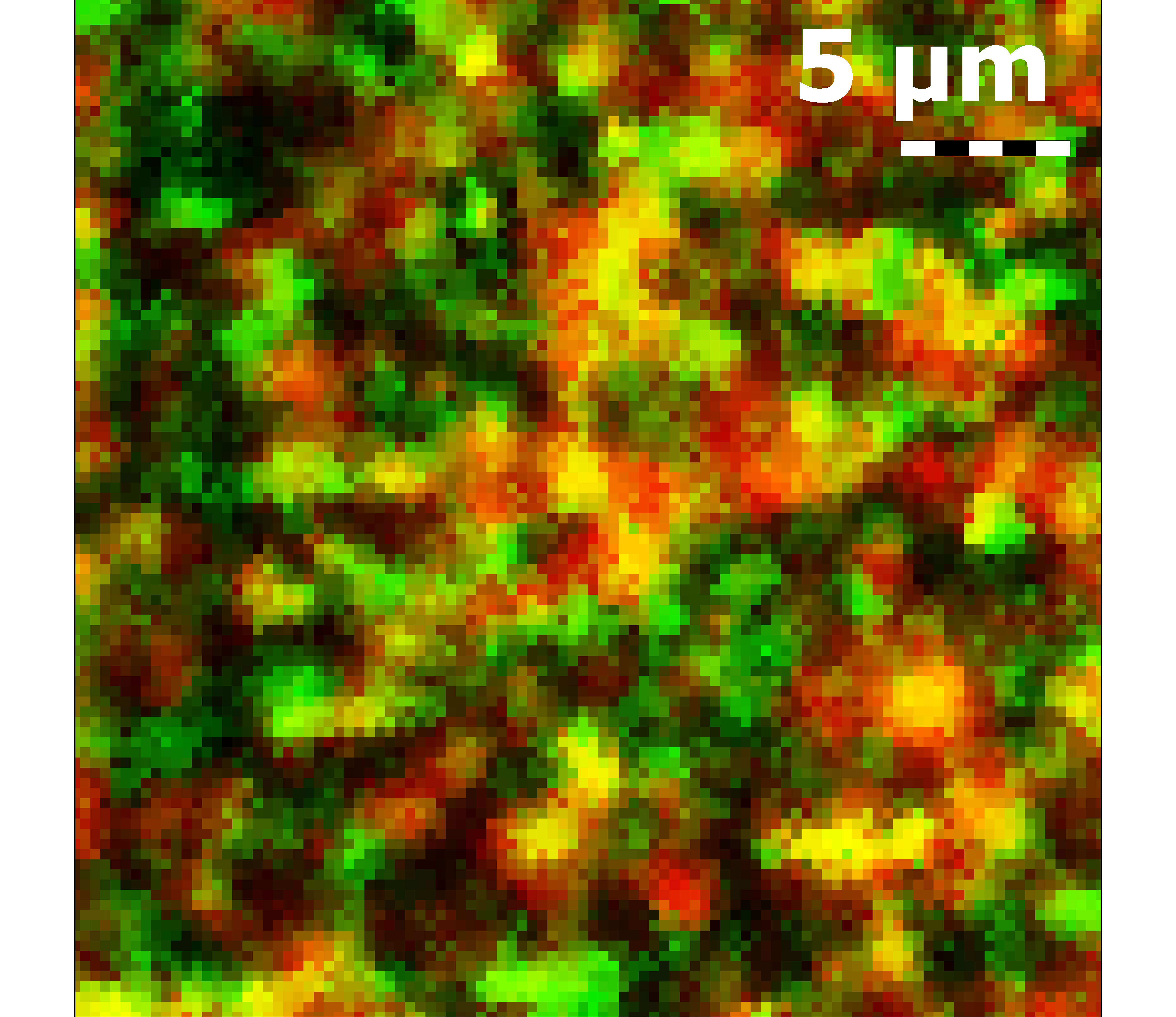}
 \caption{\textbf{Composite rendering of the integrated intensities of the 134 (red) and 157~cm$^{-1}_{}$ (green) Raman bands of the spatially resolved micro Raman measurements of the 30-minute sample.} The 134 (red) and 157 cm$^{-1}_{}$ (green) Raman bands are indicative of the GaSe and Ga$_2$Se$_3$ phases, respectively. While the overall dominating phase is GaSe, the Ga$_2$Se$_3$ is especially high at the edges of regions with strong GaSe signals.}
\label{figGa2Se3}
\end{figure}

\begin{figure}[h]
 \centering\includegraphics[width=0.9\textwidth]{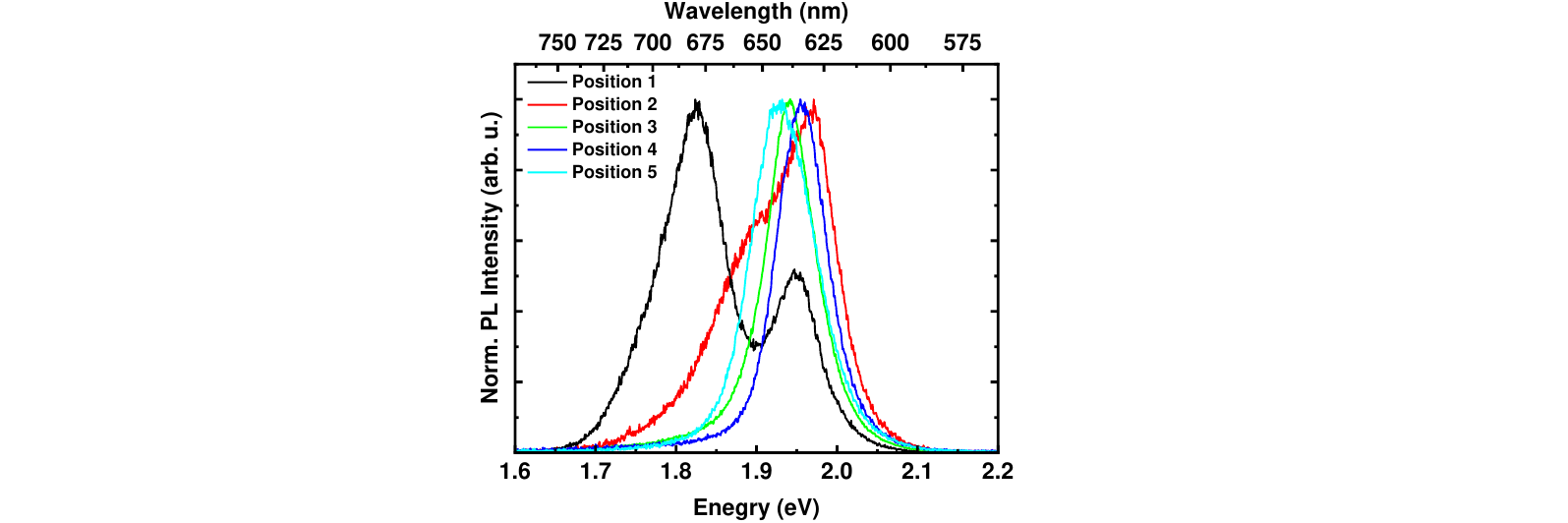}
 \caption{\textbf{Room-temperature PL line scan:} Line scan along the Raman profile in Figure~3. The spectra were normalized for better comparison. The amplitude of the highest intensity spectrum (position 3) is 12 times higher than for the least intense PL spectrum (position 3).}
\label{SI-PL-Profile}
\end{figure}

\begin{figure}[h]
 \centering\includegraphics[width=0.9\textwidth]{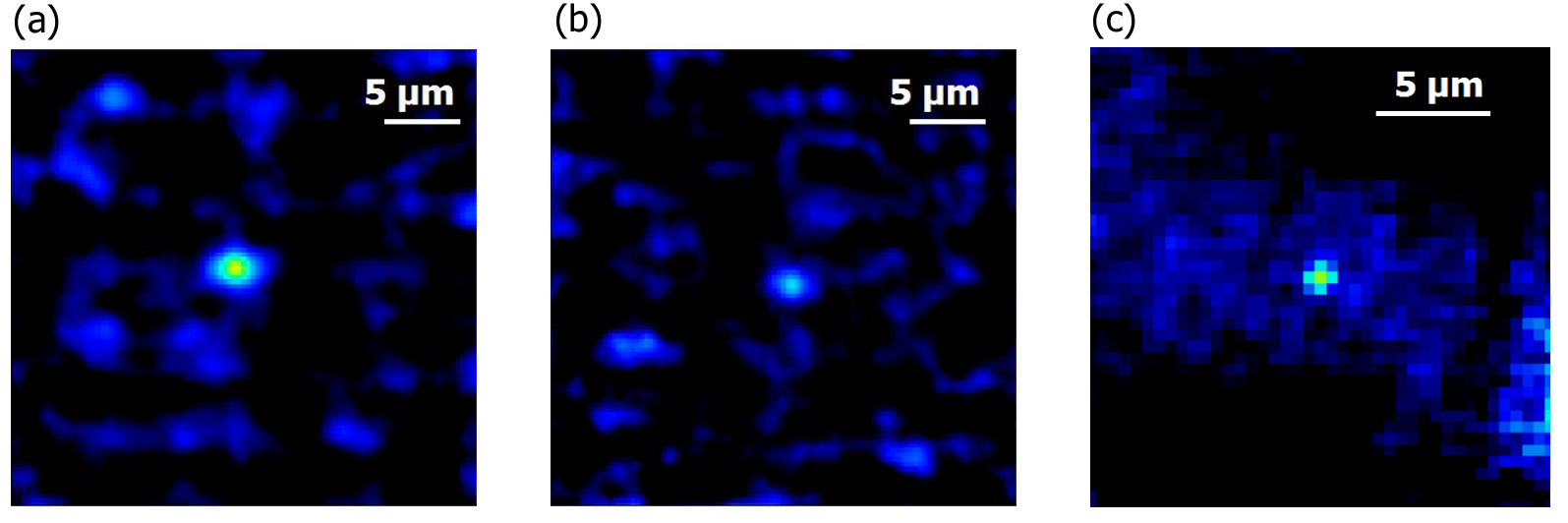}
 \caption{\textbf{Correlation maps of PL and Raman maps:} \textbf{(a)} Correlation map of the 3 minute grown GaSe sample. \textbf{(b)} Correlation map for the 135~cm$^{-1}_{}$ GaSe band of the 30 minute grown sample. \textbf{(c)} Correlation map for the 157~cm$^{-1}_{}$ of Ga$_2$Se$_3$ of the 30 minute grown GaSe sample. The z-scale for all correlation maps ranges from 0 (black) to 1 (red).}
\label{fig1PL}
\end{figure}


\bigskip

\newpage
\bibliography{sn-bibliography}